\documentclass[runningheads]{llncs}
\usepackage[latin9]{inputenc}
\usepackage{amsmath}
\newcommand{\tabincell}[2]{\begin{tabular}{@{}#1@{}}#2\end{tabular}}  
\usepackage{graphicx}
\usepackage[noend]{algpseudocode}
\usepackage{algorithmicx,algorithm}
\usepackage{cite}
\usepackage{multirow}
\usepackage{amsmath}  
\usepackage{float}
\usepackage{booktabs} 
\usepackage[noend]{algpseudocode}
\usepackage{algorithmicx,algorithm}
\usepackage{cite}
\usepackage{multirow}
\usepackage{amsmath}  
\usepackage{float}
\usepackage{booktabs} 

\usepackage{graphicx}
\usepackage{subfigure}
\usepackage{bm}
\usepackage{lineno}
\usepackage{url}
\usepackage{tikz}
\usepackage{verbatim}
\usepackage{amsfonts}
\usetikzlibrary{positioning}
\usepackage{calc}
\usepackage[colorlinks,linkcolor=black]{hyperref}
\usepackage{subfigure}
\usepackage{bm}
\usepackage{lineno}
\usepackage{url}
\usepackage{tikz}
\usetikzlibrary{positioning}
\usepackage{calc}
\usepackage{diagbox}
\usepackage{url}
\usepackage{verbatim}

\usepackage{graphicx}
\begin{document}
\title{IStego100K: Large-scale Image Steganalysis Dataset}
%
%

\author{Zhongliang Yang\inst{1} \and
Ke Wang\inst{1} \and
Sai Ma\inst{2} \and
Yongfeng Huang\inst{1} \and
Xiangui Kang\inst{3} \and
Xianfeng Zhao\inst{2}
}
%
%

\institute{Beijing National Research Center for Information Science and Technology, Tsinghua University, Beijing, 100084 \\
\email{\{yangzl15, k-w17\}@mails.tsinghua.edu.cn}\\
\email{yfhuang@tsinghua.edu.cn}\\
\and
State Key Laboratory of Information Security, Institute of Information Engineering, Chinese Academy of Sciences, Beijing 100093 \\
\email{masai, zhaoxianfeng}@iie.ac.cn
\and
Guangdong Key Lab of Information Security, Sun Yat-sen University, Guangzhou\\
\email{isskxg@mail.sysu.edu.cn}}

\maketitle              
\begin{abstract}

In order to promote the rapid development of image steganalysis technology, in this paper, we construct and release a multivariable large-scale image steganalysis dataset called IStego100K. It contains 208,104 images with the same size of 1024*1024. Among them, 200,000 images (100,000 cover-stego image pairs) are divided as the training set and the remaining 8,104 as testing set. In addition, we hope that IStego100K can help researchers further explore the development of universal image steganalysis algorithms, so we try to reduce limits on the images in IStego100K. For each image in IStego100K, the quality factors is randomly set in the range of 75-95, the steganographic algorithm is randomly selected from three well-known steganographic algorithms, which are J-uniward, nsF5 and UERD, and the embedding rate is also randomly set to be a value of 0.1-0.4. In addition, considering the possible mismatch between training samples and test samples in real environment, we add a test set (DS-Test) whose source of samples are different from the training set. We hope that this test set can help to evaluate the robustness of steganalysis algorithms. We tested the performance of some latest steganalysis algorithms on IStego100K, with specific results and analysis details in the experimental part. We hope that the IStego100K dataset will further promote the development of universal image steganalysis technology\footnote{The description of IStego100K and instructions for use can be found here: \url{https://github.com/YangzlTHU/IStego100K}}.

\end{abstract}
\section{Introduction}

Concealment system, together with encryption system and privacy system, is classified into three basic information security systems by Claude E. Shannon \cite{shannon1949communication}. Among them, the latter two security systems mainly guarantee the security of information content, but they may expose the existence and importance of information while protecting it. But for concealment system, it mainly protects the information from the perspective of behavioral security, hiding the existence of information and communication behavior, thus ensuring the security of important information. Due to its powerful information hiding ability, concealment system plays an important role in protecting the privacy and security in cyberspace.

There are various media forms of carrier that can be used for information hiding, including image\cite{fridrich2009steganography,chen2019defining}, audio\cite{yang2017sudoku,yang2018aag}, text\cite{yang2019rnn,yang2018rits,yang2018automatically} and so on\cite{johnson2008detection}. Among them, image has the characteristics of large information capacity, which has become a widely studied and used steganographic carrier in recent years. However, while protecting the security of information, these concealment systems may also be used by criminals and transmit some malicious information, thus bringing potential risks to cyberspace security \cite{theohary2011terrorist}. Therefore, studying and developing effective steganalysis techniques becomes an increasingly promising and challenging task.

For a concealment system, we can usually model it as follows. Suppose there is a carrier space $\mathcal{C}$, a key space $\mathcal{K}$, and a secret information space $\mathcal{M}$. Alice chooses a secret information $m$ from the secret space $\mathcal{M}$, under the guidance of the secret key $k\in \mathcal{K}$, uses the steganographic algorithm $f()$ to embed $m$ into a carrier $c\in \mathcal{C}$ and form the steganographic carrier $s$, that is:

\begin{equation}
Emb: \mathcal{C} \times \mathcal{K} \times \mathcal{M} \to \mathcal{S}, f(c,k,m) = s.
\end{equation}

\noindent Generally speaking, once we insert additional information into the carrier, it will inevitably lead to changes in the distribution of some features of the carriers. In order to ensure the security of the steganographic system, the steganographic algorithm $f()$ chosen by Alice should minimize the statistical differences between the carriers before and after steganography, that is:

\begin{equation}
d_f(P_{\mathcal{C}},P_{\mathcal{S}}) \leq \varepsilon.
\end{equation}

Steganalysis technology is the countermeasure technology of steganography. Its main purpose is to detect whether covert information is contained in the information carrier being transmitted in cyberspace. It can help identify potential network attacks in cyberspace and maintain cyberspace security. Any steganalysis can be described by a map $\mathit{F}: \mathbb{R}^d \rightarrow \{0,1\}$, where $\mathit{F} = 0$ means that $x$ is detected as cover, while $\mathit{F} = 1$ means that $x$ is detected as stego. Therefore, steganalysis researchers usually construct a variety of corresponding statistical features, and based on these features to find the differences in the statistical distribution between cover and stego carriers \cite{holub2014low,song2015steganalysis,boroumand2019deep,xu2016structural,wu2018deep,boroumand2019deep,yang2019ts,yang2019fast,yang2019real}.  

This paper is motivated in three aspects. Firstly, in order to achieve higher performance steganalysis technology, researchers usually need to analyze the statistical distribution differences between a large number of normal samples and steganographic samples \cite{holub2014low,song2015steganalysis,boroumand2019deep}. Especially with the development of deep learning technology, some image steganalysis methods based on deep neural network have a growing demand for data \cite{xu2016structural,wu2018deep,boroumand2019deep}. However, existing steganalysis datasets, such as the widely used BOSS dataset\cite{bas2011break}, are small in scale (10,000 images for training and 1,000 for testing), it may cause the model to ignore potential subtle differences in statistical feature distributions. Secondly, at present, many image steganalysis methods have strong pertinence. They are usually aimed at one specific steganalysis algorithm. This may lead to some steganalysis algorithms giving very good results in detecting a particular steganalysis algorithm, but might fail in detecting other steganography techniques. In order to help realize the universal steganalysis algorithm and make it more practical, we need a more diverse and universal steganalysis dataset. Thirdly, current steganalysis models are usually trained and tested on images from the same source. But in reality, it is difficult to have such perfect condition. We want to know whether the existing steganalysis models can still maintain good performance when training samples and test samples come from different image sources.

In order to promote the development of image steganalysis technology, especially the progress of universal image steganalysis technology, in this paper, we construct and release a large-scale image steganalysis dataset called IStego100K. For the first motivation, we collected 100,000 cover-stego image pairs with the same size of 1024*1024 to construct the training set. For the second motivation, each steganographic image in IStego100K is randomly embedded with three widely used image steganography (J-uniward\cite{holub2014universal}, nsF5\cite{fridrich2007statistically} and UERD\cite{guo2015using}) with a random embedding rate (bit per non-zero AC-DCT coefficient (bpnzac): 0.1-0.4). For the third motivation, we constructed two test sets. The first test set (SS-Test) contains 8,104 images from the same source as the training set. The second test set (DS-Test) contains 11,809 images from different sources of the training set. We also choose some of the latest and widely used image steganalysis models to train and test their performance on IStego100K. The experimental results are shown in details in the experimental section. We hope that IStego100K will further advance the development of image steganalysis.

In the remainder of this paper, Section II introduces related image steganalysis datasets. Section III introduces the detailed information of the IStego100K dataset, including data collection and preprocessing, information embedding algorithms. The Following part, Section IV, describes the steganalysis benchmarks we use and their performance on IStego100K dataset. Finally, conclusions are drawn in Section V.

\section{Related Dataset}

BOSS dataset \cite{bas2011break} is currently the most widely used image steganalysis dataset. It contains two databases of images, which are BOSSBase for training and BOSSRank for testing. BOSS dataset has greatly promoted the development of image steganalysis in previous years. However, with the advancement of technology, this dataset currently shows increasingly limitations. 

\begin{table}[!tp]
\renewcommand\arraystretch{1.4}
\caption{\label{tab:1}The main characteristics of BOSS and IStego100K.}
\centering
\setlength{\tabcolsep}{3mm}{
\begin{tabular}{c|c|c|c|c|c}
\toprule[1.5pt]
\multirow{2}{*}{Dataset} &\multicolumn{2}{c|}{BOSS} &\multicolumn{3}{c}{IStego100K}\\
\cline{2-6}
&BOSSBase &BOSSRank &Train &SS-Test &DS-Test\\
\hline
Number &10,000 &1,000 &200,000 &8,104 &11,809\\
\hline
Size &\multicolumn{2}{c|}{512*512} &\multicolumn{3}{c}{1024*1024}\\
\hline
Image style &\multicolumn{2}{c|}{grayscale} &\multicolumn{3}{c}{color}\\
\hline
bpnzac &\multicolumn{2}{c|}{0.4} &\multicolumn{3}{c}{0.1-0.4}\\
\hline
Steganography &\multicolumn{2}{c|}{HUGO} &\multicolumn{3}{c}{J-uniward, nsF5, UERD}\\
\bottomrule[1.5pt] 
\end{tabular}}

\end{table}

Firstly, on the scale of the dataset, BOSSBase contains 10,000 grayscale images with the same size of 512*512, and BOSSRank database contains 1,000 512*512 grayscale images. However, IStego100K contains 200,000 images for training (100,000 cover-stego image pairs), each of which is a 1024*1024 color image. The core of the stegaalysis is to find the statistical distribution difference between normal carriers and steganographic carriers through the analysis model. In general, the more samples, the more helpful for the model to discover the statistical distribution differences between the carrier features.

Secondly, steganographic images in BOSS datasets are embedded using a single steganographic algorithm HUGO \cite{bas2011break}, which hides messages into least significant bits of grayscale images represented in the spatial domain. However, a single steganographic algorithm can only bring differences in the statistical distribution of samples in a limited way. In complex real-world environments, the steganography algorithms used by Alice may be varied. It is often difficult for the detector to know which steganographic algorithm is used for a sample that may contain covert information. We hope to further promote the development of universal image steganalysis technology, so that the steganalysis model can have certain detection capabilities for a variety of steganographic methods. Therefore, we set up a variety of randomness settings for the steganographic samples in IStego100K. For example, image quality factor, image steganography algorithm and embedding rate are all set in a certain dynamic range for steganographic images in IStego100K.

Thirdly, in the real environment, the source-mismatch of training samples and test samples is a very important problem. In reality, it is a very realistic and challenging problem to train and detect sample source inconsistencies. Because in reality, it's hard for Eve to know the source of the steganographic samples Alice and Bob are transmitting, and it's equally difficult to get a large number of training samples from the same source. In fact, this requires that image steganalysis algorithms have strong robustness and can still have high steganalysis ability for different source image samples. We believe that in order to achieve more practical and general steganalysis algorithm, the problem of sample source mismatch is worth considering. Therefore, different from the BOSS dataset, we present two test sets from different sources, one from the same source as the training sample (SS-Test) and the other from different sources (DS-Test).

In order to compare the IStego100K and BOSS more intuitively, we present the main characteristics of the two datasets in Table 1.

\section{The Construction of IStego100K}

In this section, we will introduce in details of the construction process of IStego100K, including source image collection, image preprocessing and information hiding. Finally, we give the overall distribution characteristics of IStego100K.

\subsection{Source Image Collection}

All of the training images in IStego100K were crawled from Unsplash\footnote{\url{https://unsplash.com/}}, a copyright-free photography website\footnote{\url{https://unsplash.com/license}}. We first used the API provided by Unsplash website to randomly crawl a large number of high quality photographic images. From these original images, we then selected pictures whose shortest edge is greater than 1024 and whose quality factor is higher than 95. At the same time, we also filter some images with similar content and single scene artificially. Finally, we obtained 108,104 original images. In addition, in order to explore the problem of image source mismatch, we have built another test set. We collected daily photos taken by more than 30 people using their mobile phones (without private information and they all agreed to make them public for research). After manually deleting some images that did not meet the requirements, we collected a total of 11,809 images.

\subsection{Image Preprocessing}

For image steganalysis, there are many factors that can affect the final detection results, such as steganographic algorithm, image size, image quality factor (QF), steganography embedding rate, etc. To construct a universal dataset for image steganalysis, in IStego100K, we only unified the image size to be 1024*1024, and the other three factors are randomly set within a certain range. For image size, we firstly cut images into square according to the length of the short edge. Secondly, we resized the clipped images into 1024*1024. For image quality factor, we randomly adjusted the quality factor (QF) for the images obtained from Unsplash to be $\{75,80,85,90,95\}$. And we maintain the QF distribution of images obtained from the phone unchanged.

\subsection{Information Embedding}

In order to construct a general and practical dataset for image steganalysis, we choose a variety of widely used steganographic algorithms, which are J-uniward\cite{holub2014universal}, nsF5\cite{fridrich2007statistically} and UERD\cite{guo2015using}, to embed covert information into samples of IStego100K. We first randomly selected 100,000 images from the original images in IStego100K as the training set, and the remaining 8,104 images as the test set. In the information embedding process, we randomly selected one of the three steganographic algorithms and used them to embed the random bits stream into all the images in the training set and the random half of the test set (both SS-Test set and DS-Test set). For each steganographic image, the embedding rate was randomly set to be 0.1, 0.2, 0.3, and 0.4.

\begin{figure*}[ht]
\centering

\subfigure[Original Image]{
\begin{minipage}[t]{0.4\linewidth}
\centering
\includegraphics[width=\linewidth]{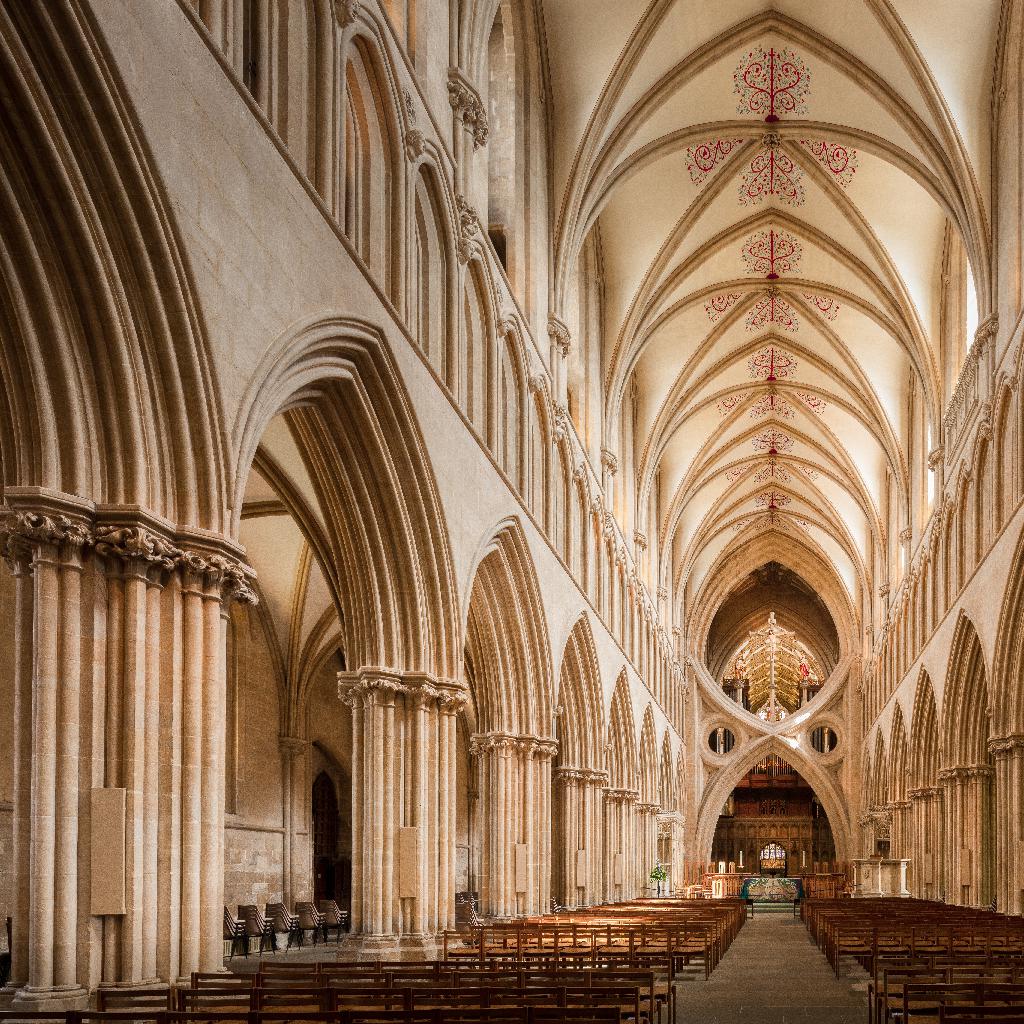}
\end{minipage}%
}%
\subfigure[J-uniward (ER = 0.2)]{
\begin{minipage}[t]{0.4\linewidth}
\centering
\includegraphics[width=\linewidth]{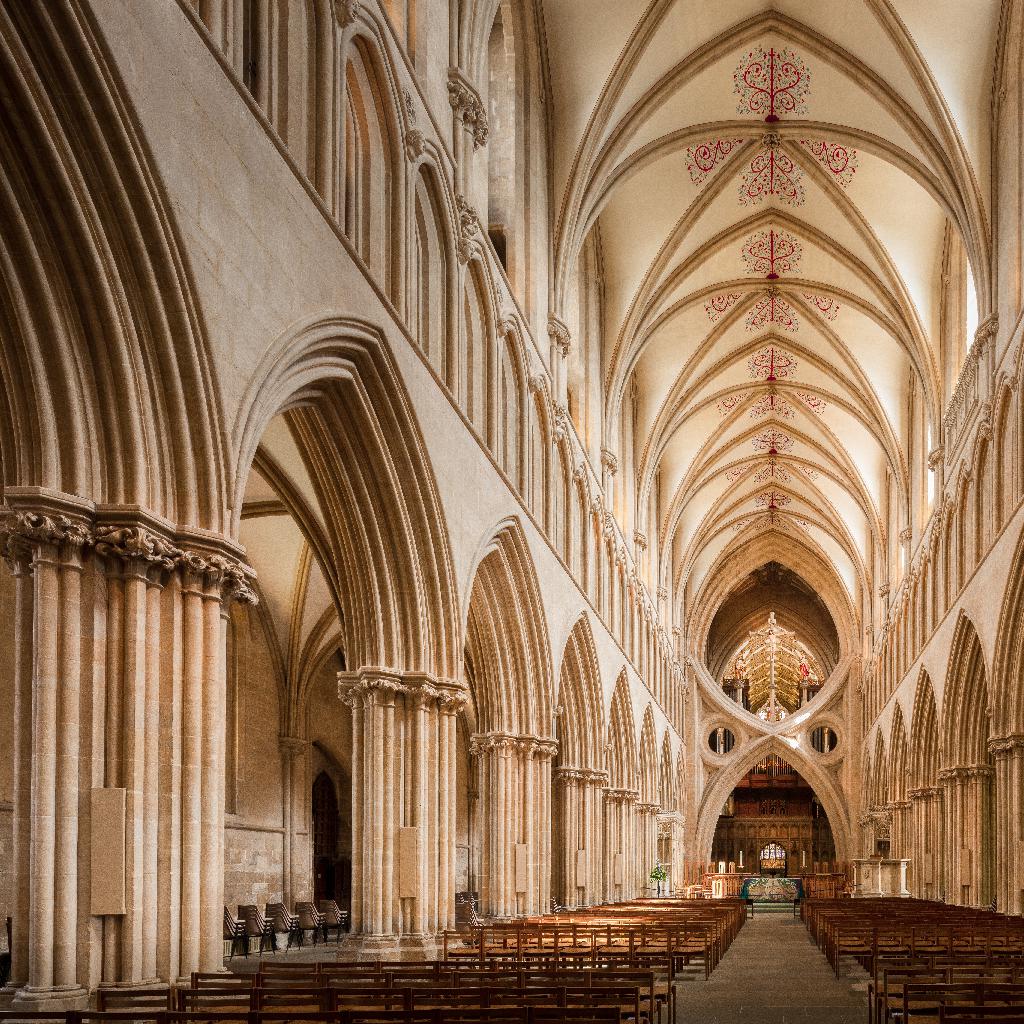}
\end{minipage}%
}%
\quad

\subfigure[nsF5 (ER = 0.2)]{
\begin{minipage}[t]{0.4\linewidth}
\centering
\includegraphics[width=\linewidth]{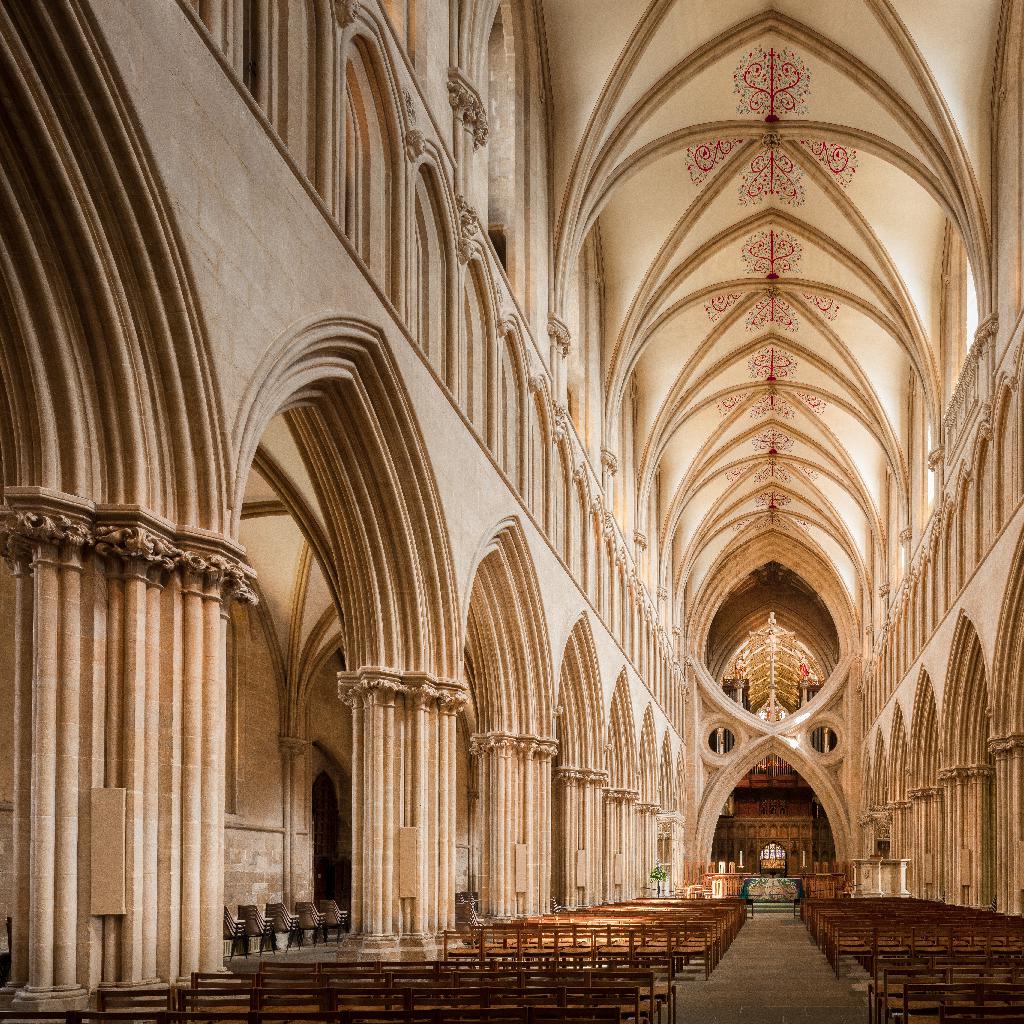}
\end{minipage}%
}%
\subfigure[UERD (ER = 0.2)]{
\begin{minipage}[t]{0.4\linewidth}
\centering
\includegraphics[width=\linewidth]{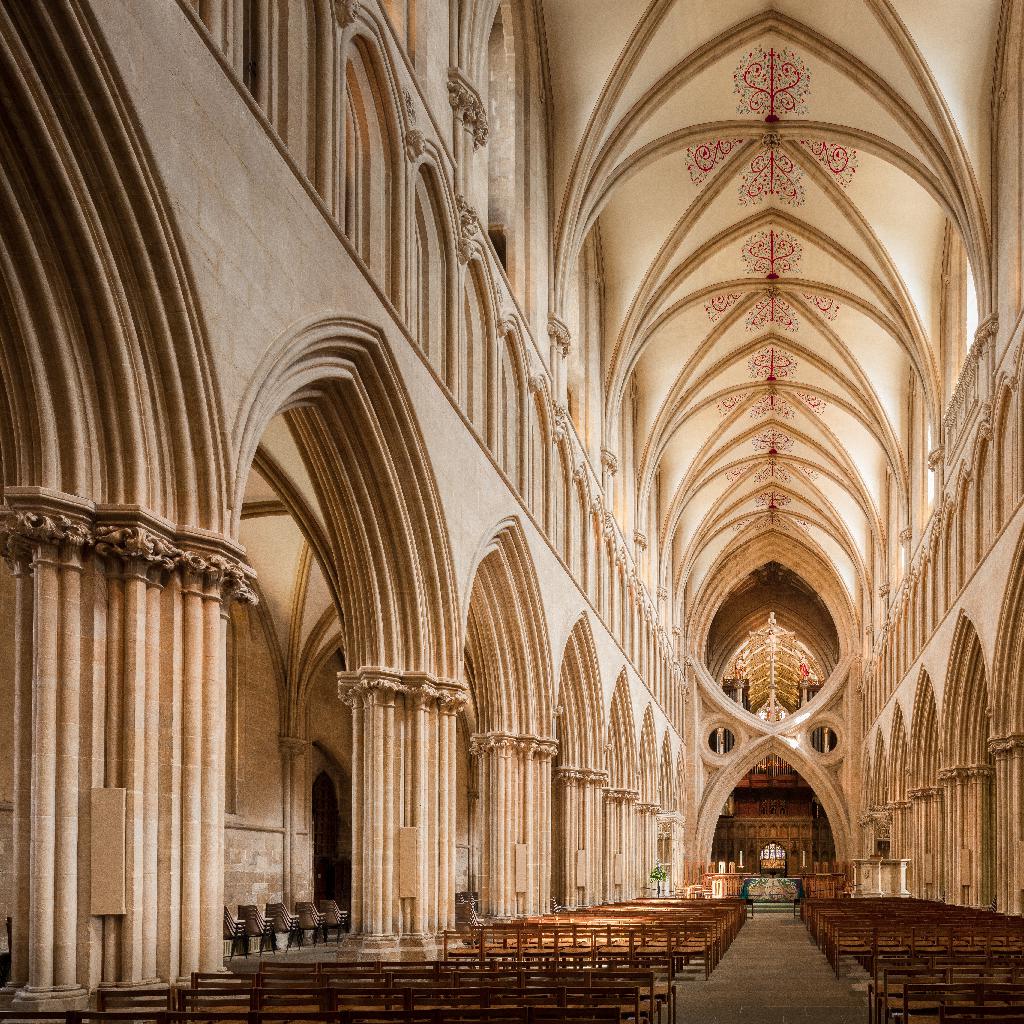}
\end{minipage}%
}%
\centering
\caption{Pictures that are embedded random bitstream by different steganographic algorithms at the same payload (bpnzac = 0.2).}
\end{figure*}

\subsection{Overall details of IStego100K}

After these above operations, the overall characteristics of IStego100K are shown in Table 2. Figure 1 shows the case when the same image is embedded by different steganographic algorithms at the same payload (bpnzac = 0.2). From the examples in Figure 1, we find that it is very difficult to distinguish the normal image from the steganographic images visually.

\begin{table}[!tp]

\centering
\caption{\label{tab:2}The overall characteristics of IStego100K.}
\resizebox{\textwidth}{16mm}{
\begin{tabular}{c|c|c|c}
\toprule[1.5pt]
IStego100K &Training Set &SS-Test Set &DS-Test Set\\
\hline
Image Number$\ $(Cover:Stego) &100,000$\ $:$\ $100,000 &4,052$\ $:$\ $4,052 &5,904$\ $:$\ $5,905\\
\hline
\tabincell{c}{Number for different steganography\\(Uerd$\ $:$\ $nsf5$\ $:$\ $j-uniward)} &33,416$\ $:$\ $33,404$\ $:$\ $33,180 &1,347$\ $:$\ $1,325$\ $:$\ $1,380 &1,969$\ $:$\ $1,981$\ $:$\ $1,955\\
\hline
\tabincell{c}{Number for different payloads\\(0.1$\ $:$\ $0.2$\ $:$\ $0.3$\ $:$\ $0.4)} &25,077$\ $:$\ $24,878$\ $:$\ $25,251$\ $:$\ $24,794 &1,047$\ $:$\ $984$\ $:$\ $1,045$\ $:$\ $976 &1,484$\ $:$\ $1,500$\ $:$\ $1,451$\ $:$\ $1,470\\
\hline
\tabincell{c}{Steganographic images of different QF\\(75$\ $:$\ $80$\ $:$\ $85$\ $:$\ $90$\ $:$\ $95)} &10,058$\ $:$\ $9,925$\ $:$\ $9,979$\ $:$\ $10,032$\ $:$\ $10,006 &803$\ $:$\ $806$\ $:$\ $820$\ $:$\ $798$\ $:$\ $825 &95.369$\pm$1.664\\
\bottomrule[1.5pt] 
\end{tabular}}
\end{table}

\section{Experimental}

\subsection{Benchmark Methods and Evaluation Metrics}

To evaluate the difficulty of IStego100K and provide benchmark results for researchers who subsequently use this dataset, we tested four latest and widely used image steganalysis methods on this proposed dataset, which are DCTR\cite{holub2014low}, GFR\cite{song2015steganalysis}, XuNet\cite{xu2017deep} and SRNet\cite{boroumand2019deep}. DCTR\cite{holub2014low} extracts the first-order statistics of quantized noise residuals obtained from the inputted image using 64 kernels of the discrete cosine transform (DCT) as features for steganalysis. GFR\cite{song2015steganalysis} extracts features based on 2-dimensional (2D) Gabor filters, which have certain optimal joint localization properties in the spatial domain and in the spatial frequency domain and can describe the image texture features from different scales and orientations, therefore it can detect the changes of statistical feature distribution before and after steganography. XuNet\cite{xu2017deep} and SRNet\cite{boroumand2019deep} and based on convolutional neural networks (CNN), for which, XuNet\cite{xu2017deep} contains a 20-layer CNN and SRNet\cite{boroumand2019deep} designed a deep residual architecture to minimize the use of heuristics and extract features, finally these features are sent to classifiers for steganlysis.

We use several evaluation indicators commonly used in classification tasks to evaluate the performance of our model, which are precision (P), recall (R), F1-score (F1) and accuracy (Acc). The conceptions and formulas are described as follows:

\begin{itemize}

\item Accuracy measures the proportion of true results (both true positives and true negatives) among the total number of cases examined

\begin{equation}
Accuracy = \frac{TP + TN}{TP + FN + FP + TN}.
\end{equation}

\item Precision measures the proportion of positive samples in the classified samples.

\begin{equation}
Precision = \frac{TP}{TP + FP}.
\end{equation}

\item Recall measures the proportion of positives that are correctly identified as such.

\begin{equation}
Recall = \frac{TP}{TP + FN}.
\end{equation}

\item F1-score is a measure of a test's accuracy. It considers both the precision and the recall of the test. The F1 score is the harmonic average of the precision and recall, where an F1 score reaches its best value at 1 and worst at 0.

\begin{equation}
F1-score = \frac{2 \times Precision \times Recall}{Precision + Recall}.
\end{equation}

\end{itemize}

\noindent TP (True Positive) represents the number of positive samples that are predicted to be positive by the model, FP (False Positive) indicates the number of negative samples predicted to be positive, FN (False Negative) illustrates the number of positive samples predicted to be negative and TN (True Negative) represents the number of negative samples predicted to be negative. All these indicators are the higher the better.

\subsection{Detection Results of Benchmark Methods}

We first used the training set in IStego100K to train various steganalysis models and then used both SS-Test set and DS-Test set for testing. Table 3 records the test performance of each steganalysis model on both test set. In the process of model training, we are surprised to find that the two steganalysis methods based on neural network, which are XuNet\cite{xu2017deep} and SRNet\cite{boroumand2019deep}, are hardly to converge on IStego100K. This may be caused by various reasons. To run these two models, we downloaded their training codes from \url{https://github.com/GuanshuoXu/caffe_deep_learning_for_steganalysis} and \url{http://dde.binghamton.edu/download/} respectively. We adopted the default training parameters, and then trained them in the environment of GTX1080TI and CUDA8.0. The small GPU memory (about 11G) may limited the performance of the model. Another main reason we thought is the diversity of samples in IStego100K, including multi-steganography, multi-quality factors, multi-embedding rates. These results at least indicate that although neural network technology and neural network-based image steganalysis models\cite{xu2017deep,boroumand2019deep} have developed rapidly in recent years, they still face enormous challenges in the face of more complex real-world environments. Although the neural network-based image steganalysis model can achieve better results than the manual feature-based steganalysis model \cite{holub2014low,song2015steganalysis} in some specific scenarios, there is still much room for improvement in the practicality and generality of the model.

\begin{table}[ht]
\renewcommand\arraystretch{1.3}
\centering
\caption{\label{tab:3}The overall performance of each benchmark methods.}
\setlength{\tabcolsep}{5mm}{
\begin{tabular}{c|c|c|c|c|c}
\toprule[1.5pt]

Dataset &Methods &Acc(\%) &P(\%) &R(\%) &F1(\%)\\
\hline
\multirow{3}{*}{SS-Test} &DCTR\cite{holub2014low} &71.34 &79.72 &57.23 &66.63\\
\cline{2-6}
                            &GFR\cite{song2015steganalysis} &66.26 &69.58 &57.97 &63.25\\
\cline{2-6}
                            &XuNet\cite{xu2017deep} &\multicolumn{4}{c}{Not Convergent}\\                            
\cline{2-6}
                            &SRNet\cite{boroumand2019deep} &\multicolumn{4}{c}{Not Convergent}\\
\hline
\hline
\multirow{3}{*}{DS-Test} &DCTR\cite{holub2014low} &56.95 &55.50 &70.11 &61.95\\
\cline{2-6}
                            &GFR\cite{song2015steganalysis} &59.12 &61.61 &48.42 &54.22\\
\cline{2-6}
                            &XuNet\cite{xu2017deep} &\multicolumn{4}{c}{Not Convergent}\\                            
\cline{2-6}
                            &SRNet\cite{boroumand2019deep} &\multicolumn{4}{c}{Not Convergent}\\                            
\bottomrule[1.5pt] 
\end{tabular}}
\end{table}

In addition, Table 3 also compares the detection performance of DCTR\cite{holub2014low} and GFR\cite{holub2014low} on the two test sets. Firstly, we noticed that the detection results of both DCTR and GFR on SS-Test is better than that on DS-Test. This result is in line with our expectations, since after all, the samples in DS-Test do not come from the same source as those in the training set. But we are also glad to see that these two steganalysis algorithms still have certain detection ability even in the case of source mis-match. This results reflect the robustness of these two steganalysis algorithms to some extent.

\begin{table}[H]
\renewcommand\arraystretch{1.3}
\centering
\caption{\label{tab:4}The detection performance of each benchmark methods for different steganography algorithms in IStego100K.}
\setlength{\tabcolsep}{2.5mm}{
\begin{tabular}{c|c|c|c|c|c|c}
\toprule[1.5pt]
Test Set &Steganalysis &Steganography &Acc(\%) &P(\%) &R(\%) &F1(\%)\\
\hline
\multirow{6}{*}{SS-Test} &\multirow{3}{*}{DCTR\cite{holub2014low}} &UERD\cite{guo2015using} &71.77 &79.75 &58.36 &67.40\\
\cline{3-7}
                      & &nsF5\cite{fridrich2007statistically} &84.44 &85.10 &83.51 &84.30\\
\cline{3-7}
                      & &J-uniward\cite{holub2014universal} &57.73 &67.58 &29.71 &41.27\\
\cline{2-7}
& \multirow{3}{*}{GFR\cite{song2015steganalysis}} &UERD\cite{guo2015using} &68.47 &71.34 &61.75 &66.20\\
\cline{3-7}
                      & &nsF5\cite{fridrich2007statistically} &71.61 &72.72 &69.18 &70.91\\
\cline{3-7}
                      & &J-uniward\cite{holub2014universal} &58.81 &62.91 &42.92 &51.02\\             
\hline
\hline
\multirow{6}{*}{DS-Test} &\multirow{3}{*}{DCTR\cite{holub2014low}} &UERD\cite{guo2015using} &53.96 &53.35 &63.06 &57.80\\
\cline{3-7}
                      & &nsF5\cite{fridrich2007statistically} &62.28 &60.56 &87.59 &71.61\\
\cline{3-7}
                      & &J-uniward\cite{holub2014universal} &51.67 &51.43 &59.83 &55.31\\
\cline{2-7}
& \multirow{3}{*}{GFR\cite{song2015steganalysis}} &UERD\cite{guo2015using} &56.05 &58.40 &42.09 &48.92\\
\cline{3-7}
                      & &nsF5\cite{fridrich2007statistically} &67.24 &68.21 &64.58 &66.35\\
\cline{3-7}
                      & &J-uniward\cite{holub2014universal} &54.59 &56.62 &39.26 &46.37\\
\bottomrule[1.5pt] 
\end{tabular}}
\end{table}

On the basis of Table 3, we have made a more detailed analysis of the detection results on test sets. We analyzed the impact of different steganographic algorithms on the detection results. We calculated the test results of different steganalysis methods on the test set for each steganographic algorithm. The results are shown in Table 4. 

From the results in Table 4, we can find that, firstly, when these three steganographic algorithms are mixed together, whether using DCTR or GFR for steganalysis, J-uniward \cite{holub2014universal} seems to be the most difficult to detect, and nsF5 \cite{fridrich2007statistically} is relatively easier to detect. To some extent, it proves that the concealment of the three steganography algorithms from strong to weak seems to be J-uniward \cite{holub2014universal}, UERD\cite{guo2015using} and nsF5\cite{fridrich2007statistically}. Secondly, when we compare the detection accuracy of two steganalysis algorithms on the two test sets, we find a very interesting phenomenon: the detection accuracy of DCTR on SS-Test seems to be better than that of GFR, but on DS-Test, GFR's detection accuracy seems to be better than DCTR's. This seems to indicate that the robustness of the GFR model is better than the robustness of the DCTR.

\begin{table*}[ht]
\renewcommand\arraystretch{1.3}
\centering
\caption{\label{tab:5}The detection performance of each benchmark methods for different embedding rates in IStego100K.}
\setlength{\tabcolsep}{1.2mm}{
\begin{tabular}{c|c|c|c|c|c|c|c|c|c}
\toprule[1.5pt]
\multicolumn{2}{c}{Test Set} &\multicolumn{4}{|c}{SS-Test}  &\multicolumn{4}{|c}{DS-Test}\\
\hline
Steganalysis &Payload &Acc(\%) &P(\%) &R(\%) &F1(\%) &Acc(\%) &P(\%) &R(\%) &F1(\%)\\
\hline
\multirow{4}{*}{DCTR\cite{holub2014low}} &0.1 &58.55 &67.84 &32.51 &43.96 &52.86 &52.42 &61.90 &56.77\\
\cline{2-10}
                      &0.2 &71.43 &80.19 &56.90 &66.57 &56.21 &54.99 &68.40 &60.97\\
\cline{2-10}
                      &0.3 &76.30 &82.22 &67.11 &73.90 &58.56 &56.53 &74.11 &64.13\\
\cline{2-10}
                      &0.4 &79.55 &83.74 &73.35 &78.20 &60.17 &57.72 &76.05 &65.63\\                      
\hline
\hline
\multirow{4}{*}{GFR\cite{song2015steganalysis}} &0.1 &55.87 &59.40 &37.10 &45.67 &52.29 &53.42 &35.79 &42.86\\
\cline{2-10}
                      &0.2 &63.51 &67.98 &51.08 &58.33 &56.66 &58.87 &44.19 &50.49\\
\cline{2-10}
                      &0.3 &70.83 &72.04 &67.89 &69.95 &62.15 &64.65 &53.65 &58.63\\
\cline{2-10}
                      &0.4 &75.71 &74.89 &76.75 &72.05 &65.40 &67.18 &60.22 &63.51\\             
\bottomrule[1.5pt] 
\end{tabular}}
\end{table*}

We further analyzed the impact of different embedding rates on the test results. We calculate the detection performance of each steganalysis method for images with different embedding rates in the test set. The results are shown in Table 5. From the results in Table 5, we can easily find a very obvious change rule, that is, as the embedding rate increases, the detection performance of each detection model is gradually improved. For example, for the DCTR algorithm \cite{holub2014low}, when the embedding rate is 0.1, the detection accuracy is only 58.55\%. When the embedding rate is increased to 0.4, the detection is also improved to 79.55\%. This trend can be explained by Formula (2). Embedding additional information in the original image carrier is equivalent to introducing noise into the original signal, which will inevitably change the statistical distribution characteristics of the original signal carrier. The higher the embedding rate, the more extra information is embedded, which will cause this statistical distribution to become larger and therefore easier to be detected. In Table 5, we found the same phenomenon as in Table 4. That is to say, from the detection accuracy, the detection accuracy of DCTR on SS-Test is higher than that of GFR, but it turns to the opposite on DS-Test.

\begin{table}[ht]
\renewcommand\arraystretch{1.3}
\centering
\caption{\label{tab:6}The detection performance of each benchmark methods for different quality factors in IStego100K.}

\setlength{\tabcolsep}{5.5mm}{
\begin{tabular}{c|c|c|c|c|c}
\toprule[1.5pt]
\multicolumn{2}{c|}{Test Set} &\multicolumn{4}{c}{SS-Test}\\
\hline
Steganalysis &QF &Acc(\%) &P(\%) &R(\%) &F1(\%)\\
\hline
\multirow{5}{*}{DCTR\cite{holub2014low}} &75 &75.23 &85.63 &60.64 &71.00\\
\cline{2-6}
                      &80 &71.50 &86.48 &61.56 &71.82\\
\cline{2-6}
                      &85 &74.09 &84.34 &59.18 &69.55\\
\cline{2-6}
                      &90 &69.04 &76.09 &55.54 &64.21\\    
\cline{2-6}
                      &95 &62.12 &66.41 &49.05 &56.43\\                                         
\hline
\hline
\multirow{5}{*}{GFR\cite{song2015steganalysis}} &75 &70.08 &75.06 &60.15 &66.78\\
\cline{2-6}
                      &80 &69.91 &74.98 &59.75 &66.50\\
\cline{2-6}
                      &85 &68.42 &71.54 &61.17 &65.95\\
\cline{2-6}
                      &90 &64.67 &67.02 &57.75 &62.04\\ 
\cline{2-6}
                      &95 &58.30 &59.76 &50.82 &54.93\\                                   
\bottomrule[1.5pt] 
\end{tabular}}
\end{table}

Further more, we also want to know how the image quality factors affect steganalysis performance. Therefore, we also calculated the detection accuracy of different detection algorithms for different quality factor images in the test sets. The results are shown in Table 6. From the results in Table 6, we can see that as the image quality factor increases, the detection accuracy of various detection algorithms gradually decreases. This seems to indicate that the higher the image quality factor within a certain range, the harder it is to detect a steganographic image. Finally, Figure 2 shows the ROC curves of these two steganography algorithms on IStego100K in different situations.

\begin{figure}[ht]
\centering
\includegraphics[width=\linewidth]{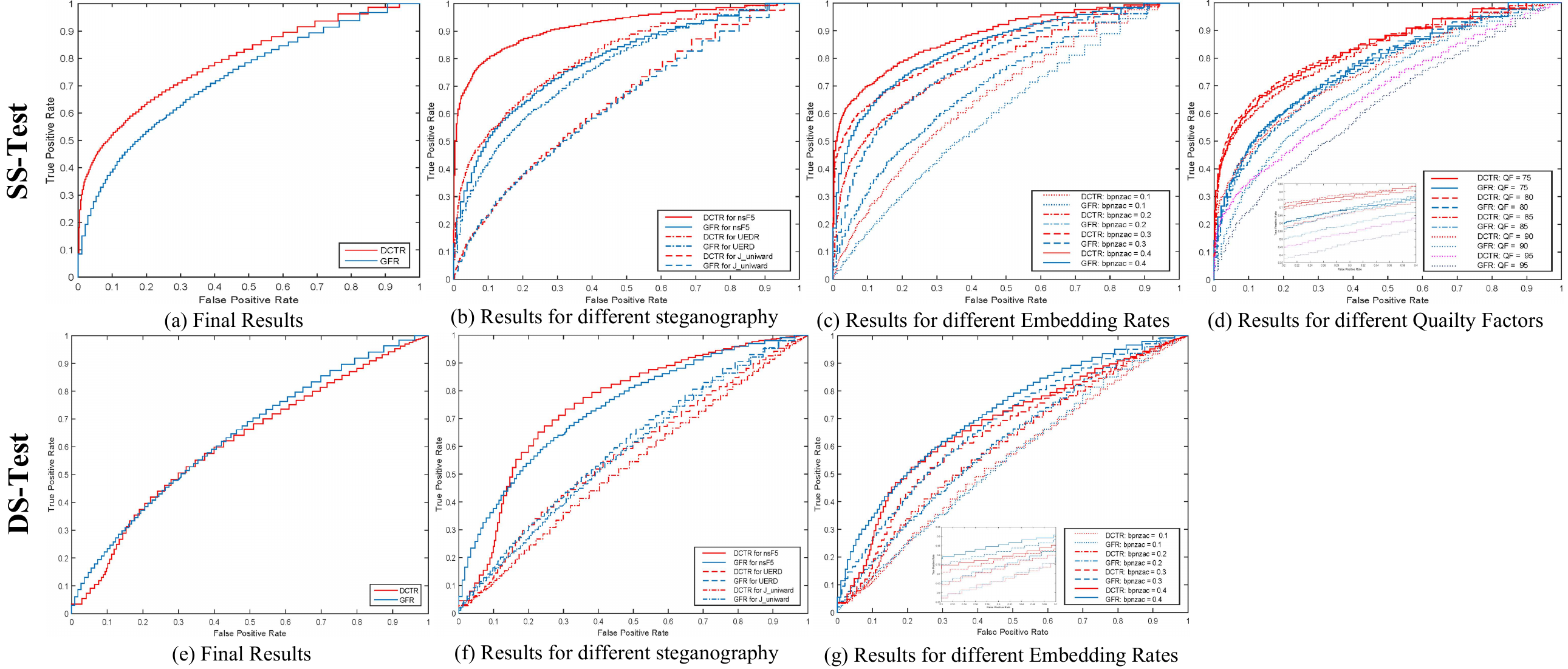}
\caption{The ROC curves of these two steganography algorithms on IStego100K in different situations.}
\label{fig:1}
\end{figure}

\section{Conclusion}

In this paper, we construct and release a large-scale image steganalysis dataset called IStego100K. It contains 208,104 images with the same size of 1024*1024, of which 200,000 images (100,000 cover-stego image pairs) construct the training set and the remaining 8,104 are as testing sets. Each steganographic image is randomly steganized with three widely used image steganography (J-uniward\cite{holub2014universal}, NSF5\cite{fridrich2007statistically} and UERD\cite{guo2015using}) with a random embedding rate (0.1-0.4). At the same time, we also choose some latest steganalysis algorithms to test IStego100K dataset. These results show some interesting phenomena. Firstly, although image analysis techniques based on convolutional neural networks have been greatly developed in recent years, and there have also appeared more and more image steganalysis techniques based on CNN. However, our detection results show that when facing with more general detection scenarios, these methods seem to still have great limitations. Secondly, the results of Table 3, 4 and 5 show that the detection performance of existing steganalysis methods will be greatly affected when facing different source detection samples from training samples. This further encourages researchers to explore more general steganalysis models for more realistic scenarios. Thirdly, the results of Tables 5 and 6 show that the image quality factor and embedding rate can significantly affect the detection performance. Generally speaking, increasing the embedding rate and reducing the quality factor in a certain range will be more helpful for steganalysis. We hope that this paper will serve as a reference guide for researchers to facilitate the design and implementation of better image steganalysis method.

\section*{Acknowledgment}
   
This work was supported in part by the National Key Research and Development Program of China under Grant SQ2018YGX210002 and the National Natural Science Foundation of China (No.U1536207, No.U1705261 and No.U1636113).


%
%
%
%
\bibliographystyle{IEEEtran}
\bibliography{IEEEexample}
\end{document}